\documentclass[fp,twocolumn]{jpsj3}
\usepackage{txfonts}
\usepackage{graphicx}

\title{Statistical Mechanics Calculations Using Variational Autoregressive Networks and Quantum Annealing}

\author{Yuta Tamura$^1$ and Masayuki Ohzeki$^{1,2,3}$}
\inst{$^1$Graduate School of Information Sciences, Tohoku University, Sendai 980-8564, Japan \\
$^2$Department of Physics, Tokyo Institute of Technology, Meguro-ku, Tokyo 152-8551, Japan \\
$^3$ Sigma-i Co., Ltd., Minato-ku, Tokyo 108-0075, Japan} 

\abst{In statistical mechanics, computing the partition function is generally difficult. An approximation method using a variational autoregressive network (VAN) has been proposed recently. This approach offers the advantage of directly calculating the generation probabilities while obtaining a significantly large number of samples. The present study introduces a novel approximation method that employs samples derived from quantum annealing machines in conjunction with VAN, which are empirically assumed to adhere to the Gibbs--Boltzmann distribution. When applied to the finite-size Sherrington--Kirkpatrick  model, the proposed method demonstrates enhanced accuracy compared to the traditional VAN approach and other approximate methods, such as the widely utilized naive mean-field.}

\begin{document}
\maketitle

\section{Introduction}

Statistical mechanics investigates the microscopic components such as atoms and molecules, utilizing statistical methods to characterize their macroscopic behaviors. For instance, the analysis of magnetic materials involves considering spin as a microscopic component to discuss material properties based on their interactions. However, as the size of these microscopic systems increases, the components become more complex, necessitating rigorous calculations that demand extensive time and computational resources. Consequently, numerous approximate computational and sampling methods have been developed in this discipline \cite{NMF, NMF2, Bethe, mcmc, MH, Dwave1,Dwave2,Dwave3}.

Among these, the naive mean-field (NMF) and Bethe approximations are prevalent \cite{NMF, NMF2, Bethe}. These methods approximate the Gibbs--Boltzmann distribution, typically challenging to compute, by preparing variational trial distributions that are directly computable. The NMF approximation neglects correlations between spins, whereas the Bethe approximation accounts for the total interaction with neighboring spins. Nevertheless, these methods exhibit limitations in strongly coupled systems and graph structures characterized by numerous loops. 

Recent advancements introduced variational autoregressive networks (VAN) \cite{autoregressive}, which is an extension of the variational NMF structure that reportedly achieved superior accuracy in obtaining physical quantities compared to both NMF and Bethe approximations. This method is independent of Markov chains and can enable the acquisition of simultaneous of multiple samples without autocorrelation and calculates upper bounds on the free energy. Despite these advantages, it encounters challenges in optimizing parameters such as the appropriate batch size and number of iterations.

An alternative approach to generating samples according to the Gibbs--Boltzmann distribution involves the Markov-Chain Monte Carlo (MCMC) method and other approximate sampling techniques \cite{mcmc, MH}. 
Recent advancements in quantum computer hardware have prompted numerous attempts to utilize this technology for sampling the Gibbs--Boltzmann distribution \cite{Dwave1, Dwave2, Dwave3}. 
Quantum annealing (QA) is a generic algorithm leveraging quantum fluctuations to address combinatorial optimization problems, has seen significant application \cite{qa1, qa2}. Specifically, the D-Wave quantum annealer developed by D-Wave Systems has facilitated commercial implementations of the QA protocol.

The introduction of the D-Wave quantum annealer has spurred efforts to identify practical applications of quantum computing. Notably, it has been employed to resolve optimization challenges in real-world contexts, including traffic flow \cite{neukart2017traffic,hussain2020optimal,inoue2021traffic, shikanai2023}, maze generation \cite{Ishikawa2023}, 
 finance \cite{rosenberg2016solving,  venturelli2019reverse}, logistics \cite{mugel_dynamic_2022}, manufacturing \cite{venturelli2016quantum, Yonaga2022, Haba2022}, preprocessing in material experiments\cite{Tanaka2023, Doi2023}, marketing \cite{nishimura2019item}, steel manufacturing \cite{Yonaga2022}, and decoding problems \cite{IdeMaximumLikelihoodChannel2020, Arai2021code}.

Furthermore, model-based Bayesian optimization has been proposed to solve intractable problems \cite{Koshikawa2021}. Benchmark tests comparing the performance of quantum annealers have been conducted in solving optimization problems \cite{Oshiyama2022}. Prior discussion have highlighted that QA in the transverse field Ising model does not equitably sample optimal solutions when multiple such solutions exist \cite{Yamamoto2020, Maruyama2021, matsuda_ground-state_2009, konz_uncertain_2019, mandra_exponentially-biased_2017, zhu_fair_2019, boixo_experimental_2013, pelofske_sampling_2021, zhang_advantages_2017}.

Due to inevitable environmental effects, the quantum annealer is deemed as a simulator for quantum many-body dynamics \cite{Bando2020, Bando2021, King2022}. Additionally, the application of quantum annealing as an optimization algorithm in machine learning has been documented\cite{neven2012qboost,khoshaman2018quantum,o2018nonnegative, Amin2018,Kumar2018,Arai2021,Sato2021,Urushibata2022,hasegawa2023}.

However, the current D-Wave quantum annealer often yields incomplete solutions due to the non-absolute zero operating environment and finite time evolution. As reported earlier, the distribution of these incomplete solution candidates by the D-Wave quantum annealer approximates the Gibbs-Boltzmann distribution \cite{qa}. This characteristic has been leveraged in applications such as Boltzmann machine learning, where the expectation of the Gibbs--Boltzmann distribution is utilized \cite{Dwave1, Dwave2, Dwave3}.
	
As an approach to statistical mechanics, variational autoregressive networks (VANs) have been extensively studied in recent years; nonetheless, challenges such as sampling efficiency, computational cost, and mode collapse persist \cite{auto1, auto2, auto3, auto4, auto5, modecollapse}. Consequently, a method integrating quantum annealing (QA) with autoregressive networks has been proposed \cite{qavan}. This technique involves training an autoregressive network with samples from the QA machine, demonstrating high efficiency in the two-dimensional spin glass model.

In this study, we propose a hybrid approach employing both QA and VAN. The innovation of this method lies in the direct use of raw samples from the QA as input to the autoregressive networks, unlike previous practices where QA samples were utilized solely for training neural networks. We compute the free energy for the Sherrington--Kirkpatrick (SK) model using two methodologies: the QA VAN, with samples from QA as input, and the MCMC VAN, with samples from MCMC as input \cite{SK}.

\section{Method}
\subsection{Variational autoregressive networks}

In the statistical mechanics model exemplified by the Ising model, the joint probability of spins $\boldsymbol{s} \in {+1, -1}^N$ adheres to the Boltzmann distribution,

\begin{equation}
    p(\boldsymbol{s}) = \frac{e^{-\beta E(\boldsymbol{s})}}{Z},
\end{equation}
where $\beta=1/T$ denotes the inverse temperature, and $Z$ indicates the partition function. 
Given the complex computation of the partition function, variational approaches like the naive mean-field (NMF) and Bethe approximations are frequently utilized. 
These methods employ an approximate distribution $q_\theta (\boldsymbol{s})$, parameterised by variational parameters $\theta$, which is simpler to compute. The objective is then to adjust $\theta$ such that $q_\theta (\boldsymbol{s})$ closely approximates the Boltzmann distribution $p(\boldsymbol{s})$.

Generally, the Kullback--Leibler (KL) divergence is employed to measure the proximity between these two distributions:

\begin{equation}
    D_{KL}(q_\theta || p) = \sum_{\boldsymbol{s}}q_\theta (\boldsymbol{s}) \ln \left(\frac{q_\theta(\boldsymbol{s})}{p(\boldsymbol{s})} \right) = \beta (F[q] - F), 
    \label{KL}
\end{equation}
where
\begin{equation}
    F[q] = \frac{1}{\beta} \sum_{\boldsymbol{s}}q_\theta (\boldsymbol{s}) \left[ \beta E(\boldsymbol{s}) + \ln q_\theta (\boldsymbol{s})\right], 
    \label{variational_free_energy}
\end{equation}
represents the variational free energy of $q_\theta (\boldsymbol{s})$ and $F=-\frac{1}{\beta}\ln Z$ indicates the free energy. 
Based on Eq. (\ref{KL}), minimizing the variational free energy $F[q]$ is equivalent to minimizing the KL divergence. 
In addition, as the KL divergence is non-zero, the variational free energy $F[q]$ provides an upper bound on the true free energy. 

Recent advancements in neural networks have facilitated the development of potent methods for representing this variational distribution $q_\theta (\boldsymbol{s})$. This VAN approach enables direct probability computation and efficient sampling. In VAN, the variational approximate distribution $q_\theta (\boldsymbol{s})$ is autoregressive, allowing each variable to be expressed as a product of conditional probabilities \cite{autoregressive, autoregressive1, autoregressive2}. 

\begin{equation}
    q_\theta (\boldsymbol{s}) = \prod_{i=1}^{N} q_\theta (s_i| s_{<i}) = \prod_{i=1}^{N} q_\theta (s_i| s_1, ..., s_{i-1}),
    \label{VAN}
\end{equation}
where N denotes the total number of spins. We note that in VAN, the distribution in Eq. (\ref{VAN}) is represented by a neural network to compute statistical mechanics in a variational manner.
The learning process in VAN is optimized using gradient-based machine learning algorithms such as Adam, which evaluate the gradient of the variational free energy as expressed in Eq. (\ref{grad F}).
\begin{equation}
    \beta \nabla_\theta F[q] = \mathbb{E}_{\boldsymbol{s} \sim q_\theta(\boldsymbol{s})} \left[ \nabla_\theta \ln q_\theta(\boldsymbol{s})\{\ln{q_\theta(\boldsymbol{s})} + \beta E(\boldsymbol{s})\}\right],
    \label{grad F}
\end{equation}
VAN is distinguished by its capability to provide an upper bound on the true free energy compared to existing frameworks such as MCMC and tensor networks \cite{tensor}, its efficiency in generating independent samples without the reliance on Markov chains, and its suitability for parallelization.

\subsection{Variational autoregressive networks with prior distribution}
\begin{figure}
\centering
 \includegraphics[width=0.9\linewidth]{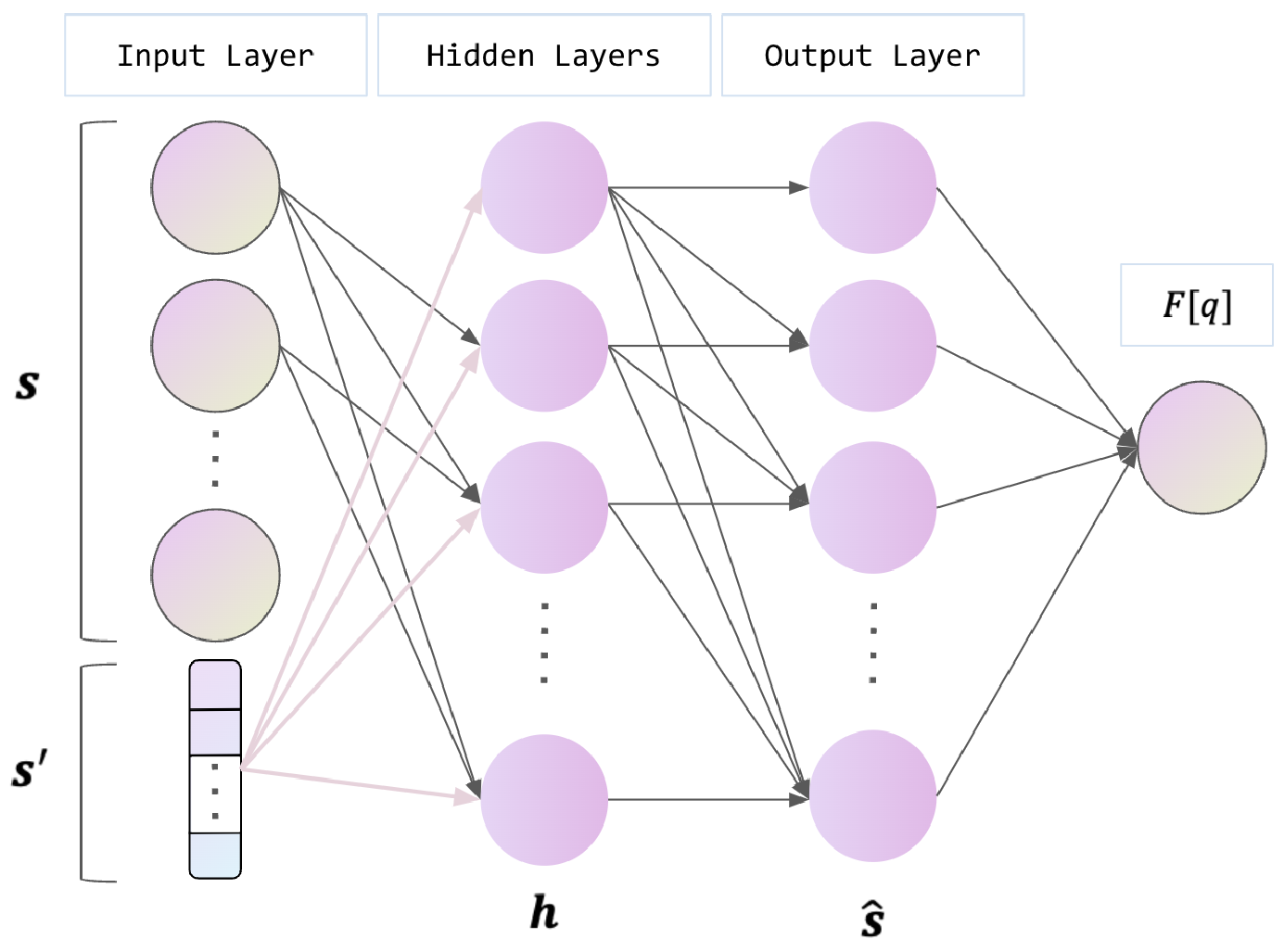} 
\caption{Architectural diagram of an autoregressive network with a prior distribution. The spin configuration $\boldsymbol{s}$ and $\boldsymbol{s}^\prime$ represent the
inputs to the network, $\hat{\boldsymbol{s}}$ indicates the output of the network. $\boldsymbol{h}$ represents the hidden layer, and this quantity is arbitrary. The variational free energy $F[q]$ is calculated by Eq. (\ref{variational_free_energy})}
\label{fig:prior_VAN}
\end{figure}

Previous studies have highlighted that VAN is a formidable approach to addressing statistical mechanics problems relative to traditional approximation methods like NMF and Bethe \cite{autoregressive}. Nonetheless, several challenges persist, including the efficiency of the sampling and learning steps, the conditions under which learning is conducted, and the domains of applicability. For instance, in the learning phase of the gradient method, opportunities still exist for optimizing the annealing schedule \cite{autoregressive}. The present method innovates by integrating VAN with a prior distribution $p(\boldsymbol{s^\prime})$ to achieve a more accurate approximation.

\begin{eqnarray}
    q_\theta(\boldsymbol{s}) &=& \sum_{\boldsymbol{s}^\prime}q_\theta(\boldsymbol{s}|\boldsymbol{s}^\prime) p(\boldsymbol{s}^\prime), \\
    q_\theta(\boldsymbol{s}|\boldsymbol{s}^\prime) &=& \prod_{i=1}^{N}q_\theta (s_i | s_1, ..., s_{i-1}, \boldsymbol{s}^\prime),
    \label{prior_VAN}
\end{eqnarray}
The approximate distribution represented by Eq. (\ref{prior_VAN}) is variably trained by deploying it through a neural network, as depicted in Fig. \ref{fig:prior_VAN}. Here, the spin $\boldsymbol{s}^\prime \in {+1, -1}^N$ is such that the prior distribution $p(\boldsymbol{s}^\prime)$ can be computed for any given oracle. For this prior distribution, we utilized samples from MCMC and QA \cite{mcmc, MH, qa}. In Fig. \ref{fig:prior_VAN}, the hidden layer $\boldsymbol{h}$ and the output layer $\boldsymbol{\hat{\boldsymbol{s}}}$ are computed as follows: 
\begin{eqnarray}
    h_i &=& \sigma \left(\sum_{i>j} W_{ij}^1 s_j + \sum_{j}W_{ij}^2 s^\prime_j \right), \label{hidden_cal}\\
    \hat{s}_i &=& \sigma \left(\sum_{i \geq j} W_{ij}^3 h_j \right), \label{output_cal}  
\end{eqnarray}
where $W^1, W^2$, and $W^3$ denote the weight matrices of the neural network and constitute the learning parameters. The spin configuration $\boldsymbol{s}$ determined by the output layer $\hat{\boldsymbol{s}}$ is governed by the binomial distribution as expressed in Eq. (\ref{sample_s}). Additionally, the spin configuration $\boldsymbol{s}^\prime$ is obtained by sampling directly from the prior distribution $p(\boldsymbol{s}^\prime)$, as stated in Eq. (\ref{sample_s_prime}). 
\begin{eqnarray}
    s_i &\sim& \hat{s_i}^{\frac{1+s_i}{2}}(1-\hat{s_i})^{\frac{1-s_i}{2}}, \label{sample_s}\\
    \boldsymbol{s}^\prime &\sim& p(\boldsymbol{s}^\prime),\label{sample_s_prime} 
\end{eqnarray}

The variational free energy $F[q]$ and its gradient $\nabla_\theta F[q]$ can be calculated using Eqs. (\ref{variational_free_energy})-(\ref{prior_VAN}) by sampling similar to Eqs. (\ref{F_prior})-(\ref{grad_F_prior}). 
\begin{eqnarray}
    F[q] &=& \frac{1}{\beta} \sum_{\boldsymbol{s}}q_\theta(\boldsymbol{s})(\ln{q_\theta(\boldsymbol{s})} + \beta H(\boldsymbol{s})) \\
    &=& \frac{1}{\beta} \sum_{\boldsymbol{s}} \sum_{\boldsymbol{s}^\prime } q_\theta(\boldsymbol{s}, \boldsymbol{s}^\prime)\left(\ln{q_\theta(\boldsymbol{s})} + \beta H(\boldsymbol{s})\right) \label{F_prior}\\
    &=& \frac{1}{\beta}\mathbb{E}_{\boldsymbol{s}, \boldsymbol{s}^\prime \sim q_\theta (\boldsymbol{s}, \boldsymbol{s}^\prime)}\left[ \ln{q_\theta(\boldsymbol{s})} + \beta H(\boldsymbol{s}) \right], 
\end{eqnarray}
\begin{equation}
    \beta \nabla_\theta F[q] = \mathbb{E}_{\boldsymbol{s}, \boldsymbol{s}^\prime \sim q_\theta (\boldsymbol{s}, \boldsymbol{s}^\prime)}\left[\nabla_\theta \ln{q_\theta(\boldsymbol{s}|\boldsymbol{s}^\prime)}\{\ln{q_\theta(\boldsymbol{s})} + \beta H(\boldsymbol{s})\}\right],
    \label{grad_F_prior}
\end{equation}
where 
\begin{equation}
    q_\theta(\boldsymbol{s}, \boldsymbol{s}^\prime) =q_\theta(\boldsymbol{s}|\boldsymbol{s}^\prime)p(\boldsymbol{s}^\prime) \\
\end{equation}
represents the joint probability distribution of spin configuration $\boldsymbol{s}$ and $\boldsymbol{s}^\prime$ . 
If the prior distribution $p(\boldsymbol{s})$ is informed about the target distribution, a more accurate approximation can be anticipated by integrating the shape of the distribution. The computational cost of sampling with this method remains comparable to that of VAN, as it only requires an increase in the number of input variables.

\subsection{Quantum annealing}
QA is an algorithm designed to solve combinatorial optimization problems leveraging quantum effects \cite{qa1}. Several of these optimization challenges can be represented by the problem of finding the ground state of the Ising model:
\begin{equation}
    E(\boldsymbol{s}) = -\sum_{i=1}^{N} c_i s_i - \sum_{i > j} J_{ij} s_i s_j,
    \label{ising_model}
\end{equation}
where $J_{ij}$ represents the interaction between adjacent spins, and $c_i$ indicates the local longitudinal magnetic field acting on the $i$-th spin. 

Although QA machines aim to find the ground state of the Ising model as depicted in Eq. (\ref{ising_model}), their real-world performance often yields solutions that deviate from the ground state owing to an imperfect operating environment and unresolved technical challenges. However, it has been reported that the output distribution $p_\mathrm{QA}(\boldsymbol{s}^\prime)$ of their output solutions is approximate to the Gibbs--Boltzmann distribution.\cite{qa, Dwave4} 
This resemblance has spurred interest in using QA outputs to train Boltzmann machines \cite{Dwave1, Dwave2, Dwave3}. Such methodologies are promising for future advancements, as they are highly efficient and can effectively utilize samples that would otherwise be discarded in the pursuit of the ground state.

\section{Results}
To demonstrate the effectiveness of VANs with prior distribution in minimizing variational free energy, we conducted experiments on a specific instance of the Sherrington--Kirkpatrick (SK) model \cite{SK}, wherein spins are interconnected by couplings $J_{ij}$ derived from a Gaussian distribution with variance $1/N$. As prior distributions, we constructed approximate distributions using samples from QA machines (QA VAN) and MCMC (MCMC VAN). For QA VAN, the D-Wave Advantage 4.1 system was utilized for QA, and for MCMC VAN, the Metropolis--Hastings algorithm was employed \cite{mcmc, MH}. The interactions reflected those in the SK model instance under investigation. Additionally, for MCMC VAN, experiments were executed at high ($\beta=0.5$) and low ($\beta=2.0$) temperatures to assess the influence of the prior distribution. Neural networks were trained over 10,000 steps using the gradient-based optimization algorithm Adam \cite{gradient_descent}. To circumvent mode collapse \cite{modecollapse}, the training started at an inverse temperature $\beta=0$ and linearly and progressively increased to the desired value.

\begin{figure}
\centering
 \includegraphics[width=0.9\linewidth]{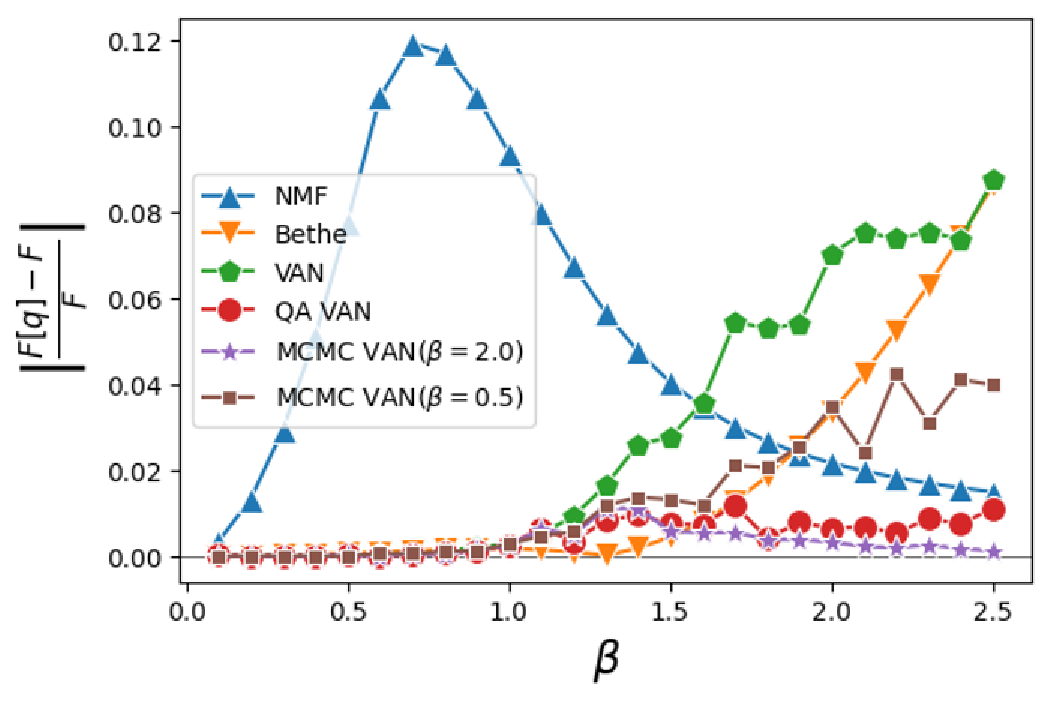} 
\caption{Relative error obtained for SK model ($N=30$). In the expectation of VANs, the calculation is performed for $N_{\mathrm{batch}}=256$.}
\label{fig:result_256}
\end{figure}

First, the results for the case where $N=30$ and $N_\mathrm{batch}=256$, which is the number of samples used to calculate the expected value in Eq. (\ref{grad_F_prior}), are displayed in Fig. \ref{fig:result_256}. For all VANs, we use the structure without hidden layers. In addition, for the relative error $\frac{F[q]-F}{F}$, $F$ denotes the exact free energy added up for all $2^{30}$ configurations. Both QA VAN and MCMC VAN ($\beta=2.0$) produced superior results compared to the conventional VAN method. At low temperatures, these methods outperformed other methods because the near-cold samples from QA and MCMC positively impacted the learning of VANs. 

The QA VAN ($\beta=0.5$) with samples from high temperatures displays significantly worse performance at low temperatures. Additionally, the better accuracy of the proposed VANs compared to naive VAN, although the number of samples used for the expected value calculation is as small as $N_{\mathrm{batch}}=256$, indicates the robustness of the proposed VANs. 

\begin{figure}
\centering
 \includegraphics[width=0.9\linewidth]{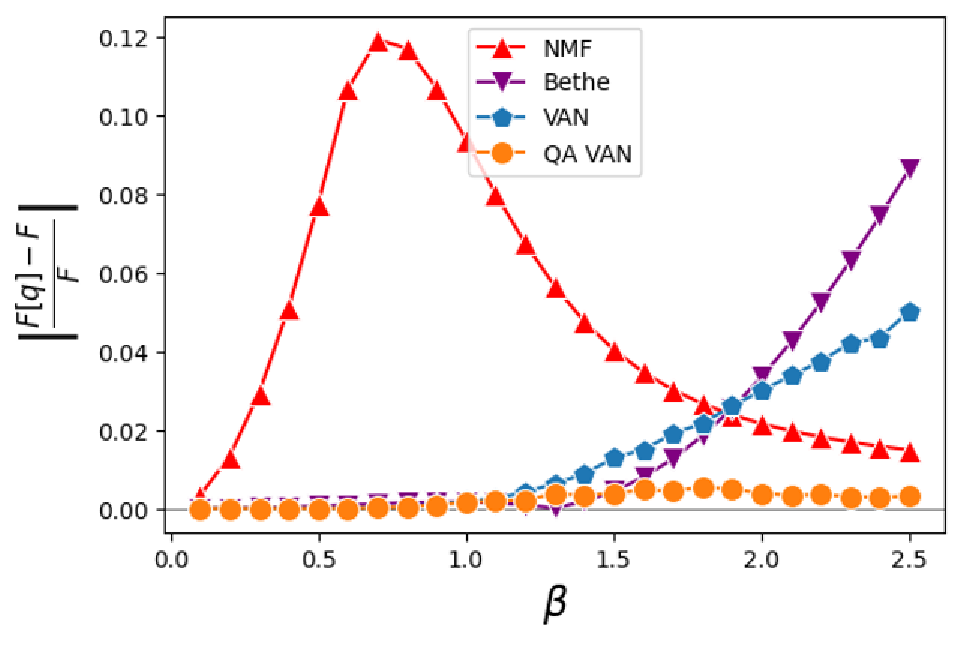} 
\caption{Relative error obtained for SK model ($N=30$). In the expectation of VANs, the calculation is performed for $N_{\mathrm{batch}}=8192$. }
\label{fig:result_8192}
\end{figure}

Figure \ref{fig:result_8192} depicts the results when a larger sample size of $N_\mathrm{batch}=8192$ is used. Figure \ref{fig:result_256} portrays that the low-temperature MCMC VAN is superior to QA VAN. Thus, in this experiment, we considered only QA VAN for comparative analyses. As observed in Fig. \ref{fig:result_8192}, the accuracy of both VAN and QA VAN can be significantly improved, and the approximation is stable and highly accurate for all inverse temperature $\beta$. 
For a small sample size, the performance of QA VAN is superior to that of all other methods in case of low temperatures. Conversely, as the temperature approaches lower temperatures, the performance of the VAN deteriorates significantly, which is apparently caused by the increase in the update range of the inverse temperature $\beta$ during the learning step. We believe that the performance can be improved by adjusting the scheduling of the inverse temperature to minimize the effect of mode decay. 

\section{Conclusion}
In this work, we elucidate the utility of VANs augmented by prior distributions. Specifically, when QA and low-temperature MCMC are utilized for prior distribution, the free energy of the SK model can be computed with exceptional accuracy across various approximation methods, including NMF and Bethe approximation. Although previous frameworks have effectively utilized incomplete samples from QA, this study broadens the potential application to diverse statistical mechanics models. Moreover, the capability to directly calculate the free energy heralds the prospect of a novel framework employing QA. Nonetheless, this method operates as a black box owing to the inherent uncertainties associated with QA, highlighting the need for further empirical research on QA applications.

\begin{acknowledgment}

Our study received financial support from the programs for Bridging the gap between R\&D and the IDeal society (society 5.0) and Generating Economic and social value (BRIDGE) and the Cross-ministerial Strategic Innovation Promotion Program (SIP) from Cabinet Office.

\end{acknowledgment}

\bibliographystyle{jpsj}
\bibliography{citation/app, citation/nn, citation/approximate, citation/van, citation/mcmc, citation/qa}

\end{document}